\documentclass[sigconf]{acmart}
\AtBeginDocument{%
  \providecommand\BibTeX{{%
    \normalfont B\kern-0.5em{\scshape i\kern-0.25em b}\kern-0.8em\TeX}}}

\setcopyright{acmcopyright}
\copyrightyear{2018}
\acmYear{2018}
\acmDOI{XXXXXXX.XXXXXXX}

\acmConference[MSR 2024]{21st International Conference on Mining Software Repositories}{April 2024}{Lisbon, Portugal}

\usepackage{xspace}
\usepackage{tikz,pifont,pgfplots}
\usepackage{xcolor, colortbl}
\usepackage{makecell,multirow,diagbox}
\usepackage{hyperref}
\usepackage[many]{tcolorbox}
\usepackage{graphicx}%
\usepackage{adjustbox}
\usepackage{amsmath,amsfonts}%
\usepackage{mathrsfs}%
\usepackage[title]{appendix}%
\usepackage{xcolor}%
\usepackage{textcomp}%
\usepackage{manyfoot}%
\usepackage{booktabs}%
\usepackage{algorithm}%
\usepackage{algorithmicx}%
\usepackage{algpseudocode}%
\usepackage{listings}%
\usepackage[frozencache,cachedir=.]{minted}
\usetikzlibrary{positioning}
\usepackage[normalem]{ulem}

\pgfplotsset{compat=1.18}

\newcommand*{\ie}{i.e.,\@\xspace}
\newcommand*{\eg}{e.g.,\@\xspace}

\newcommand*{\tool}{CodeLL\@\xspace}

\newcommand*{\GH}{GitHub\@\xspace}
\newcommand*{\PP}{PyPI\@\xspace}
\newcommand*{\DB}{Debian\@\xspace}
\newcommand*{\GL}{GitLab\@\xspace}

\begin{document}

\title{\tool: A Lifelong Learning Dataset to Support the Co-Evolution of Data and Language Models of Code}


\author{Martin Weyssow$^\ast$, Claudio Di Sipio$^\dagger$, Davide Di Ruscio$^\dagger$, Houari Sahraoui$^\ast$}
\affiliation{%
  \institution{$^\ast$DIRO, Universit\'e de Montr\'eal}
  \country{}}
\email{martin.weyssow@umontreal.ca, sahraouh@iro.umontreal.ca}
\affiliation{%
  \institution{$^\dagger$University of l’Aquila}
  \country{}}
\email{{claudio.disipio,davide.diruscio}@univaq.it}

\renewcommand{\shortauthors}{Weyssow et al.}

\begin{abstract}
Motivated by recent work on lifelong learning applications for language models (LMs) of code, we introduce \tool, a lifelong learning dataset focused on code changes.
Our contribution addresses a notable research gap marked by the absence of a long-term temporal dimension in existing code change datasets, limiting their suitability in lifelong learning scenarios.
In contrast, our dataset aims to comprehensively capture code changes across the entire release history of open-source software repositories.
In this work, we introduce an initial version of \tool, comprising 71 machine-learning-based projects mined from Software Heritage. 
This dataset enables the extraction and in-depth analysis of code changes spanning 2,483 releases at both the method and API levels.
\tool enables researchers studying the behaviour of LMs in lifelong fine-tuning settings for learning code changes. 
Additionally, the dataset can help studying data distribution shifts within software repositories and the evolution of API usages over time.

\end{abstract}




\maketitle

\section{Introduction}
\label{sec:introduction}
The abundance of open-source software (OSS) data has played a pivotal role in advancing the field of automated software engineering. 
Lately, large-scale pre-training datasets, exemplified by The Pile~\cite{gao2020pile} and The Stack~\cite{Kocetkov2022TheStack} have enabled the emergence of large language models (LLMs) tailored for programming languages understanding and generation~\cite{chen2021evaluating, roziere2023code, nijkamp2023codegen}. 
From another perspective, datasets and benchmarks such as CodeXGlue~\cite{lu2021codexglue}, CodeSearchNet~\cite{husain2019codesearchnet}, XLCoST~\cite{zhu2022xlcost} or CodeNet~\cite{puri2021codenet} have been proven to be essential for fine-tuning language models (LMs) of code\footnote{We use ``language models'' to refer to both ``large language models'' and ``pre-trained language models''.} across diverse code-related tasks.

In contrast to most previous research in code intelligence that often views these models as static entities, our stance aligns with prior research~\cite{weyssow_usage_2023, gao_keeping_2023, weyssow2023exploring}, conceptualizing LMs of code as dynamic and adaptable. 
In essence, we contend that viewing models of code as dynamically evolving over time opens up exciting opportunities to study their behaviour in lifelong learning scenarios.
This paradigm shift underscores the necessity for mining software repositories and constructing datasets that incorporate concepts related to the long-term evolution of code and repositories.
Furthermore, it sheds light on a potential research gap in the current landscape.
First, the datasets and benchmarks mentioned previously lack a temporal dimension and are commonly utilized for pre-training or fine-tuning models at a specific point in time. 
The absence of temporality compromises their ability to capture the inherent dynamic evolution in software development and repositories.
Alternatively, code change datasets such as CodeChangeNet~\cite{lin2023cct5} and MegaDiff~\cite{monperrus2021megadiff}, only encompass code changes over two specific points in time, rendering them unsuitable for pre-training or fine-tuning LMs of code in a lifelong learning setting.
Furthermore, while prior research effectively studied LMs of code in lifelong learning settings~\cite{weyssow_usage_2023, gao_keeping_2023}, there remains an absence of datasets incorporating notions of code evolution over time. 
Consequently, in the context of lifelong learning, where LMs of code evolve as dynamic entities accumulating knowledge over time, the nature of the existing datasets poses significant limitations.

To address this gap, we present an initial version of \textbf{Code} \textbf{L}ifelong \textbf{L}earning (\tool), a  dataset conceived for supporting lifelong co-evolution of software repositories and LMs of code. 
To build \tool, we leverage Software Heritage (SWH)~\cite{Cosmo2017SoftwareHW} platform that stores, collects, and maintains software repositories from multiple sources, including GitHub, PyPi, and GitLab.
The main rationale behind this choice is that SWH enables the fingerprinting of large and reproducible datasets through queries and timestamps~\cite{lefeuvre_fingerprinting_2023}.
Consequently, it renders the extraction of our raw datasets reproducible and facilitates its extension in the future.
In the scope of this paper, we collect 71 Python repositories related to machine learning that often use fast-evolving third-party libraries~\cite{10.1145/3453478}, which constitutes an interesting scope of analysis for studying lifelong code changes.
For each repository, we extract all its associated releases and Python files, resulting in a total of 2,483 releases and 970,277 files. 
Next, we design mapping heuristics to map files, methods, and API calls between two consecutive releases to analyze how methods content and API calls evolve over the entire lifespan of the repositories.

By conducting a code change analysis of one of the mined repositories (see Section~\ref{sec:code_change_analysis}), we highlight potential research opportunities spanning from the fine-tuning of LLMs of code for learning lifelong code changes and API evolution to analyzing potential distribution shifts of the data over time.
Furthermore, the descriptive statistics of our data (see Section~\ref{sec:descriptive_statistics}) underscores the potential for building a more massive version of \tool, that could be used for continual pre-training of LLMs, by mining more repositories as incorporating 71 repositories already produces a relatively large-scale dataset.



\section{Data Collection methodology}
\label{sec:proposed}
\begin{figure}[!t]
    \centering
    \includegraphics[width=1\linewidth]{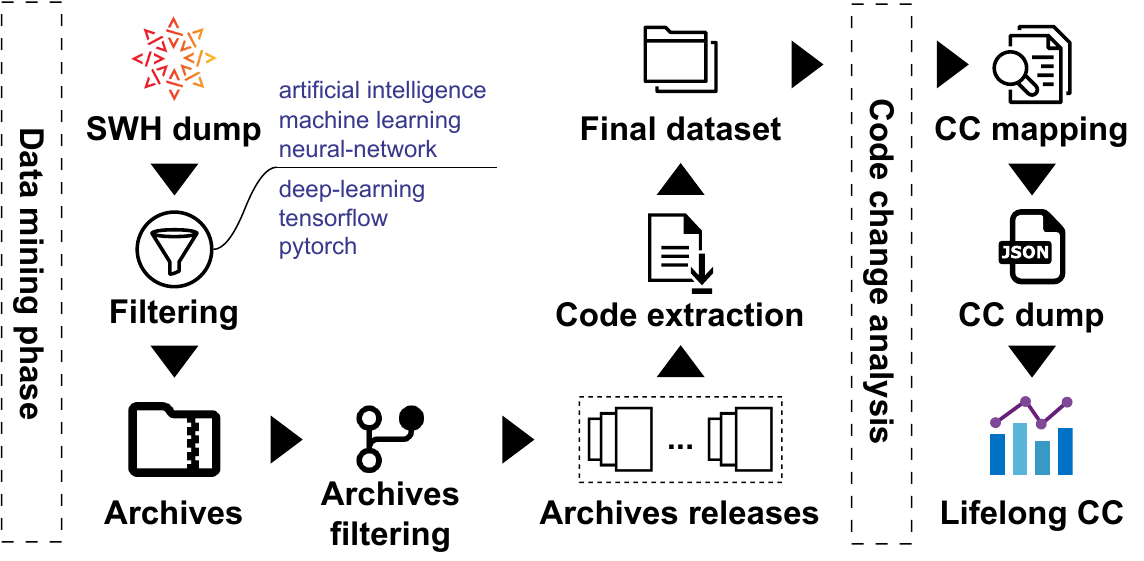}
    \vspace{-2em}
    \caption{Data collection methodology.}
    \label{fig:process}
\end{figure}

Our data collection methodology is depicted in Figure \ref{fig:process}. We describe the data mining and code change analysis phases in Section~\ref{sec:data_mining} and Section~\ref{sec:code_analyzer}, respectively.

\subsection{\tool Data Model} \label{sec:datamodel}
\begin{figure}[!t]
    \centering
    \includegraphics[width=1\linewidth]{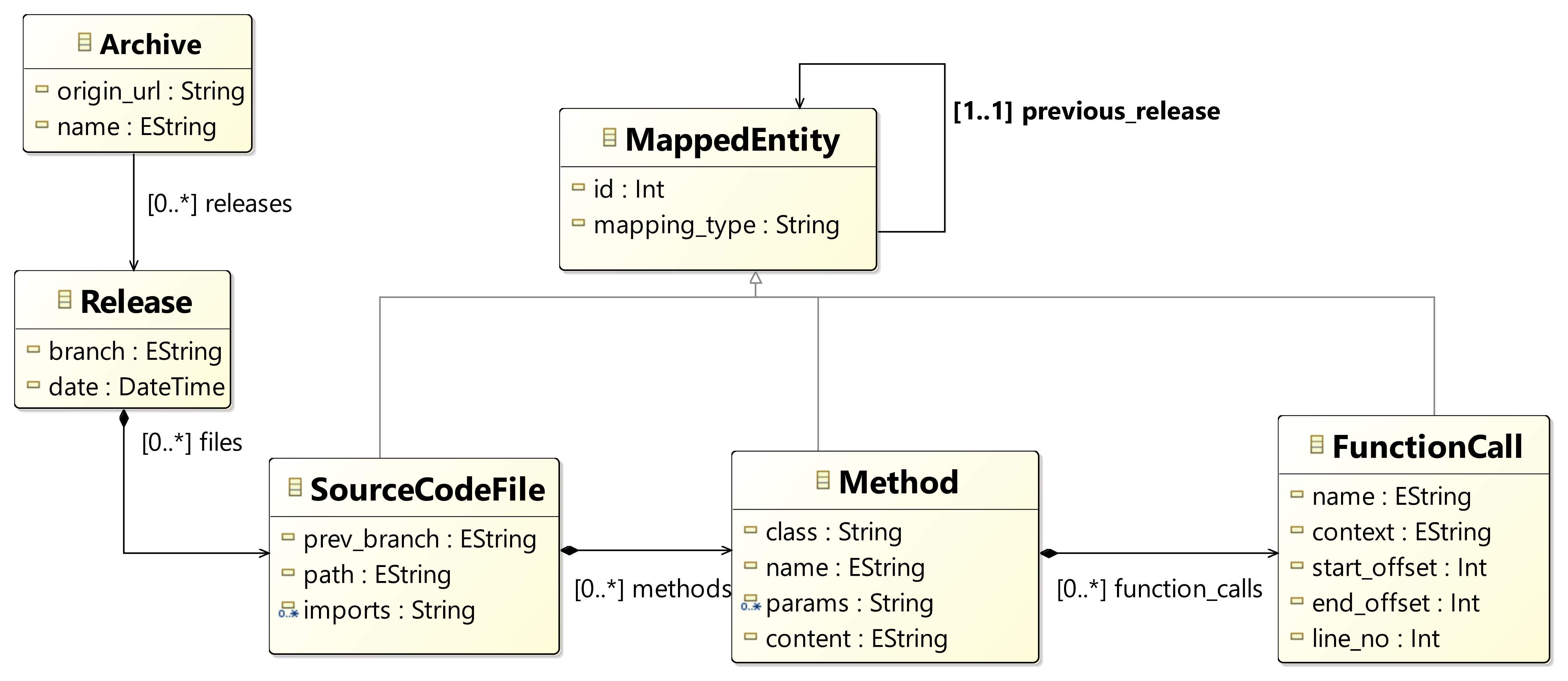}
    \caption{The \tool data model.}
    \label{fig:datamodel}
\end{figure}

Before delving into the description of the data collection methodology, we present the \tool data schema modeled as a class diagram\footnote{Interested readers can refer to \cite{lefeuvre_fingerprinting_2023} for a comprehensive presentation of the SWH general data model.} in Figure \ref{fig:datamodel}.
The diagram illustrates an \texttt{Archive} as a general software repository stored on SWH, with a unique \texttt{originURL} and multiple \texttt{releases}. 
Each \texttt{Release} is identified by a \texttt{branch} name and a \texttt{date}, \ie the timestamp of that specific release. 
A release contains a list of \texttt{SourceCodeFile} entities, each linked to its previous release through the \texttt{previous\_release} association. 
A \texttt{SourceCodeFile} is characterized by \texttt{imports} and composed of \texttt{methods}; a \texttt{Method} contains a list of \texttt{function\_calls}. It is worth noting that imports are stored as a list of strings since analysing them in depth is not in the scope of this paper. 
These entities are specialized forms of a \texttt{MappedEntity}, characterized by a unique hash ID, signifying the mapping between \texttt{SourceCodeFile}, \texttt{Method}, and \texttt{FunctionCall} originating from two distinct releases.
The \texttt{mapping\_type} attribute of \texttt{MappedEntity} denotes whether an entity is effectively mapped, added, or removed from one release to another.

\subsection{Data Mining}
\label{sec:data_mining}

\subsubsection*{SWH dump}
To collect the initial set of artifacts, we rely on the \texttt{2021-03-23-popular-3k-python} dataset available on SWH. 
The dump contains a total number of 2,197 Python artifacts originating from four different data sources, \ie 580 \GH and 135 \GL repositories, and 827 \PP and 655 \DB packages.
We leverage this dump as it encompasses only popular software repositories and packages from diverse data sources.\footnote{See \url{https://tinyurl.com/swhdataset} for more details.}



\subsubsection*{Filtering} 
From the initial dump, we exclude \DB packages due to the absence of \texttt{.py} files.
Next, we filter out repositories and packages unrelated to machine learning by relying on \GH topics\footnote{\url{https://github.com/topics}} and \PP classifiers.\footnote{\url{https://pypi.org/classifiers/}} 
As mentioned in the introduction, machine learning projects tend to use APIs that evolve at a fast pace, \eg PyTorch, which is interesting for studying the lifelong evolution of the projects. 
Specifically, we focus on artifacts labeled with terms like ``machine learning" and ``neural network", as illustrated in Figure \ref{fig:process}. 
In cases where artifacts do not align with any search term, we automatically evaluate whether they use any third-party libraries related to machine learning. 
Inclusion is then determined based on this assessment. 
This initial filtering process results in 140 archives, comprising 44 from \PP, 96 from \GH, and none from \GL. 
Given our primary objective of releasing a dataset tailored for lifelong learning studies, we focus the data mining on archive releases, \ie a version of the archive at a certain timestamp. 
We discard archives with less than two releases as it does not enable any code evolution analysis, and end up with 71 archives composed of 26 \PP packages and 45 \GH repositories.
From those archives, we extracted 2,483 releases. 
Another reason to limit our mining efforts is motivated by the potential for a single archive to contain a substantial number of releases, and the SWH API imposes restrictions on the number of requests per user. 
Nonetheless, we plan on expanding the dataset by incorporating all the artifacts contained in the \texttt{2021-03-23-popular-3k-python} SWH dump in the future.


\subsubsection*{Code extraction}
For each release, we statically analyze the \texttt{.py} files they contain. Within each file, we retrieve imports, standalone functions, and class methods. Furthermore, we extract the function and API calls within each function and method, and store their code positions, \ie start and end offsets.
Importantly, we establish implicit links between API calls and imports; \eg ``torch.mean'' is connected to ``import torch''. This information storage facilitates the application of API-related downstream tasks using \tool.

\subsection{Code Change Analysis}
\label{sec:code_analyzer}
Given two releases $r_m$ and $r_n$ of a software archive, the code change analysis involves mapping files, methods, and function/API calls from both releases. 
Note that $r_m$ and $r_n$ do not have to be consecutive releases, \ie the code change analysis can be performed between any archive releases.

\subsubsection*{Files mapping}
We start by mapping files with identical relative paths. 
To handle cases where a file in $r_m$ changes path in $r_n$, as in the case of code refactoring, we rely on the following heuristics. 
We ensure a high Levenshtein similarity ratio between both file names ($\geq$ 0.75) and verify the presence of identical functions/methods in their content.

\subsubsection*{Methods mapping} 
We map functions and methods belonging to two mapped files $f_{r_m}$ and $f_{r_n}$ if they belong to the same class (\texttt{None} if the function is standalone), have the same names and parameters. 
If a method from $f_{r_m}$ remains unmapped, we map it with a method from $f_{r_n}$ that belongs to the same class and has the same name.

\subsubsection*{Function calls mapping}
We map function and API calls belonging to two mapped methods $m_{f_{r_m}}$ and $m_{f_{r_n}}$ if they have the same code positions.
If a function/API call in $m_{f_{r_m}}$ remains unmapped, we map it with an identical function/API call in $m_{f_{r_n}}$ with the closest position.

Finally, unmapped files, methods and function/API calls $\in r_m$ are tagged as ``removed'', while unmapped elements $\in r_n$ are tagged as ``added''.

\section{Data overview}
\label{sec:overview}
In this section, we present an overview of \tool by first discussing descriptive statistics of the dataset. 
Subsequently, we conduct a code change analysis of one selected repository from our dataset.
Finally, we discuss the storage and reusability of \tool.

\subsubsection*{Descriptive statistics}
\label{sec:descriptive_statistics}

\begin{figure}
    \centering
    \includegraphics[width=.9\linewidth]{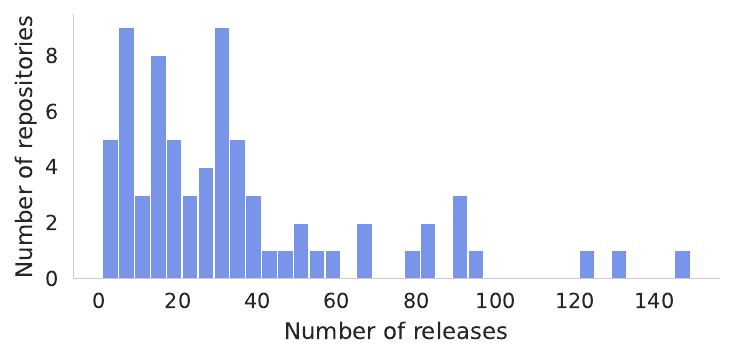}
    \vspace{-1.5em}
    \caption{Number of releases per repository.}
    \label{fig:dist_releases}
\end{figure}

\begin{figure}
    \centering
    \includegraphics[width=.9\linewidth]{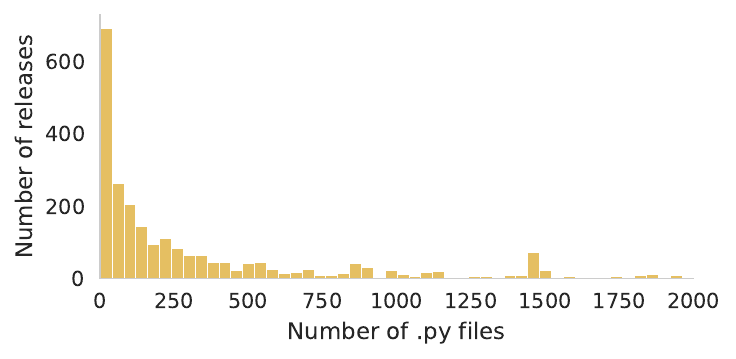}
    \vspace{-1.5em}
    \caption{Number of Python files per release.}
    \label{fig:dist_files}
\end{figure}

This initial version of \tool comprises 71 software repositories, 2,483 releases, and 970,277 Python files.
Figure~\ref{fig:dist_releases} illustrates the distribution of releases per software repository, revealing that the majority feature between 0 and 40 releases, with certain repositories exhibiting a notably higher release count.
This diversity facilitates examining software repository evolution across a spectrum of lifespans.
Furthermore, Figure~\ref{fig:dist_files} depicts the distribution of files per release, underscoring the capacity for extensive evolution analysis across a substantial number of files.

\subsubsection*{Lifelong code change analysis}
\label{sec:code_change_analysis}

\begin{figure}
    \centering
    \includegraphics[width=\linewidth]{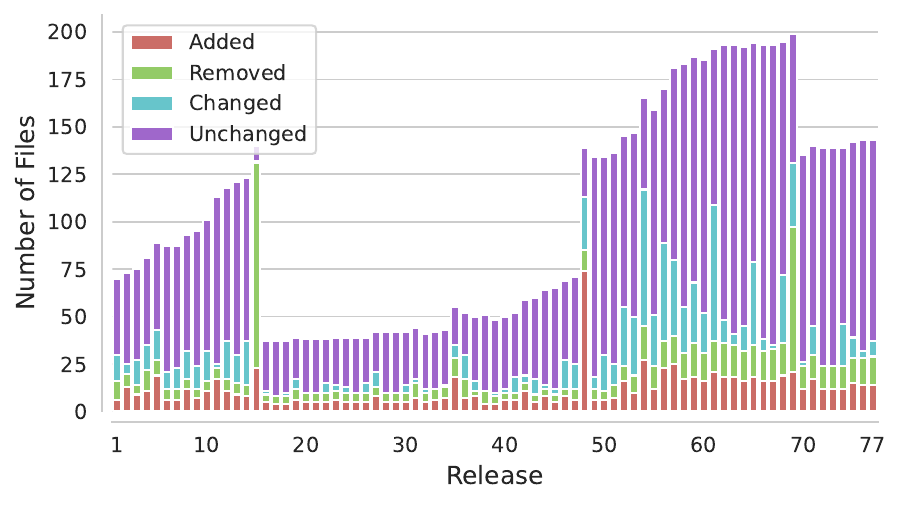}
    \vspace{-2.5em}
    \caption{Gensim project file changes.}
    \label{fig:files_changes}
\end{figure}

\begin{figure}
    \centering
    \includegraphics[width=\linewidth]{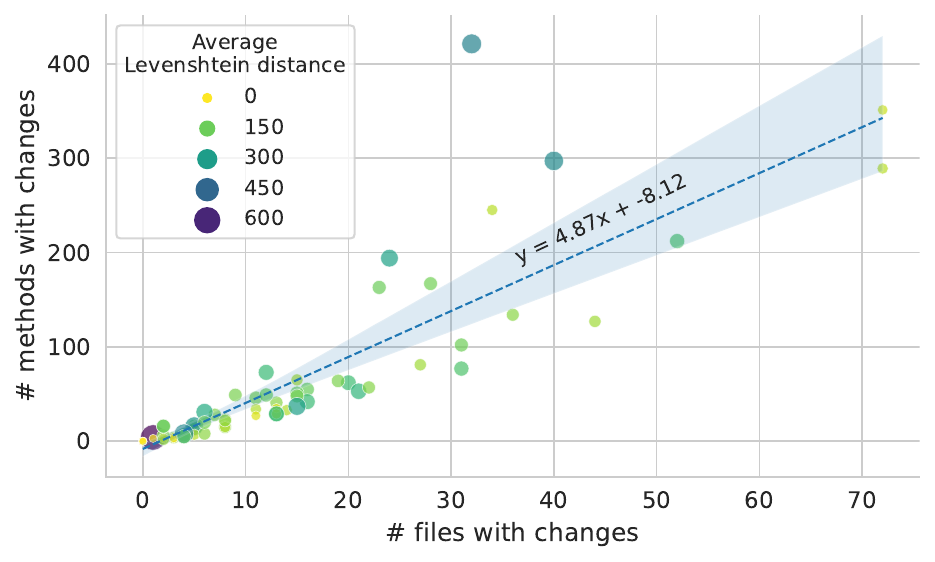}
    \vspace{-2.5em}
    \caption{Correlation between changed files and methods in each Gensim release, with Levenshtein distance computed and averaged across changed methods.}
    \label{fig:levenshtein}
\end{figure}

For illustrative reasons, we analyse the evolution of the Gensim project, which comprises 78 releases.
In Figure~\ref{fig:files_changes}, we analyse how files change throughout the releases.
This analysis helps pinpointing releases where either substantial refactoring occur, \eg at release \#15, or instances where a significant amount of files changed from one release to another, \eg between releases \#52-70.
Furthermore, Figure~\ref{fig:levenshtein} shows a linear correlation between the number of files and methods that changed for each release of Gensim.
The regression plot also depicts the average Levenshtein distance computed for each altered method, facilitating the identification of releases with substantial changes at the code level.
Overall, these analyses are pivotal in identifying distinctive patterns in the evolution of a software repository, streamlining the application of subsequent data analyses or downstream tasks to specific releases.

\subsubsection*{Storage and reusability}
\label{sec:StorageAndReusability}

\begin{listing}
\begin{minted}[
               linenos=true,
               xleftmargin=21pt,
               tabsize=2,
               fontsize=\footnotesize]{js}
{     
  "repository": str, "branch" : str, "prev_branch" : str,
  "date": str, "id": integer, "imports": List[str],
  "methods": List[{
    "id": integer, "class": str, "name": str, 
    "params": List[str], "content": str, 
    "function_calls": List[{
      "id": integer, "expression": str,
      "start_offset": integer, "end_offset": integer,
      "mapping": integer | "added" | "removed"
    },
    "mapping": integer | "added" | "removed"
  }],
  "mapping": integer | "added" | "removed"
}
\end{minted}
\caption{Description of one line of the \tool dataset.} 
\label{json-example}
\end{listing}

Our dataset and code are available online.\footnote{ \url{https://github.com/martin-wey/CodeLL-Dataset}}
We provide one \texttt{.jsonl} file per mined repository, and Listing~\ref{json-example} illustrates the structure of one line of the \texttt{.jsonl} files.
We also provide our code to mine more data from SWH, and generate the \texttt{.jsonl} files, as well as a Python notebook to compute statistics.

\section{Research opportunities}
\label{sec:opportunities}

\noindent
\textbf{Lifelong learning for code and LLMs}.
The primary objective of \tool is to enable the study of (L)LMs of code in lifelong learning settings. 
As demonstrated in previous work~\cite{weyssow_usage_2023, gao_keeping_2023}, pre-trained language models of code suffer from catastrophic forgetting, \ie they forget about past data, in scenarios where the model continuously learns on new data over time.
However, these prior studies do not consider learning from the lifelong evolution of software repositories and leverage rather synthetic datasets not encompassing any temporal dimension.
In this context, our dataset enables researchers to study how LMs of code can (1) continuously acquire knowledge over the whole lifespans of software repositories during fine-tuning, or (2) adapt to a specific software release at inference using prompting techniques.
In its current state, \tool is not suitable for pre-training LLMs, but subsequent large-scale versions of \tool might enable exploring continual pre-training of LLMs.

\noindent
\textbf{Data distribution shift}.
Analyzing \tool in terms of the evolution of the data distribution over time presents a compelling research opportunity.
As discussed in Section~\ref{sec:overview}, conducting a lifelong analysis of code changes within an archive enables the identification of software releases marked by significant code changes. 
However, it is crucial to note that releases featuring code changes may not necessarily pose challenges for LMs of code, particularly if the distribution of these changes does not deviate significantly from the training data distribution.
Consequently, investigating the dynamic shifts in release distribution allows for pinpointing instances where LMs may exhibit weak generalization.
These challenges, particularly those associated with out-of-distribution generalization and long-tailed distributions, have demonstrated a substantial impact on the effectiveness of the model~\cite{zhou2023devil, weyssow_usage_2023}.

\noindent
\textbf{API evolution}.
The problem of API evolution has been studied from different perspectives, including investigating community trends \cite{alrubaye_how_nodate,kula_developers_2018}, and performing automated migration \cite{NGUYEN2022117267,xu_meditor_2019}. 
Despite these efforts, our current understanding is that there is a gap in the literature concerning the adaptation of (L)LMs of code to either changes in API usage or in the APIs themselves over extended periods.
While prior work enhanced LLMs of code with API knowledge~\cite{patil2023gorilla, qin2023toolllm}, it remains unclear how to continuously adapt LLMs to new APIs over time in an effective manner.


\section{Limitations}
\label{sec:threats}
This first version of \tool presents some limitations related to the constrained scope of mined software repositories (see Section~\ref{sec:data_mining}) and the code change analysis (see Section~\ref{sec:code_analyzer}).
First, we limited the mining to machine learning-related archives, resulting in the mining of 2,483 releases, a time-consuming process as SWH API restricts the number of requests allowed per hour. 
We believe this initial release of \tool already enables a broad range of applications as described in Section~\ref{sec:opportunities}. 
We intend to broaden the dataset with more repositories in subsequent versions of \tool.
Additionally, the current method employed for mapping files, methods, and function calls between two releases relies on heuristics.
We believe the applied heuristics enable mapping releases elements in an accurate way.
However, we acknowledge that refining these heuristics may further improves the mappings for more specific code changes, especially at the line or API call level.




\section{Related work}
\label{sec:related}
We identified code change datasets as the most related to our contribution.
To the best of our knowledge, there is no existing dataset such as \tool facilitating lifelong learning, data distribution shift, and code change/API evolution applications over the entire lifespan of software repositories.
Existing datasets, including MegaDiff~\cite{monperrus2021megadiff}, CodRep~\cite{chen2018codrep}, or ManySStuBs4J~\cite{karampatsis2020often}, focus on Java code changes without incorporating timestamps or considering the temporal evolution of code changes.
Other prior work, exemplified by APIDiff~\cite{brito2018apidiff} and PyMIgBench~\cite{islam_pymigbench_2023}, enable detecting API changes.
Our dataset not only encompasses the analysis of API changes but also enables studying how API usages within a repository evolve throughout its lifespan.
Nonetheless, we recognize that incorporating the taxonomy of code changes from PyMIgBench~\cite{islam_pymigbench_2023} in \tool would enhance our dataset.
Finally, CodeChangeNet \cite{lin2023cct5} comprises 2M data samples with code changes ranging between 3 to 100 tokens across six languages. 
Similar to other datasets, CodeChangeNet does not encompass any notion of code changes over time, rendering it unsuitable for lifelong learning applications.
Nevertheless, our plans include expanding our tool and dataset to support multiple programming languages.

\section{Conclusion and future work}
\label{sec:conclusion}
In this paper, we introduce \tool, a lifelong learning dataset to support the co-evolution of source code data and LMs of code.
The initial version of \tool encompasses 71 Python machine learning software repositories, comprising 2,483 releases and nearly 1 million files. 
\tool records code changes at both the method and function/API call levels throughout the entire release history of code repositories, facilitating the exploration of lifelong learning applications for pre-trained language models and LLMs of code.
Our dataset opens up a wide range of future work opportunities related to lifelong learning for code, identifying distribution shifts in software repository releases, and learning code change and API evolution.
As future work, we aim to expand \tool with additional software repositories spanning diverse programming languages.
Finally, we plan to delve into downstream tasks and applications using \tool.


\bibliographystyle{ACM-Reference-Format}
\bibliography{references}


\end{document}